\documentclass[review]{elsarticle}
\usepackage{lineno,hyperref}
\usepackage{booktabs}
\usepackage{color}
\usepackage{url}
\usepackage{amsmath,amsfonts,amssymb}

\modulolinenumbers[5]

\journal{Journal of \LaTeX\ Templates}

\begin{document}

\begin{frontmatter}

\title{One System to Rule them All: a Universal Intent Recognition System for Customer Service Chatbots}

\author{Juan Camilo Vasquez-Correa$^{a}$\corref{mycorrespondingauthor}, Juan Carlos Guerrero-Sierra$^{a}$, Jose Luis Pemberty-Tamayo$^{a}$, Juan Esteban Jaramillo$^{a}$, Andres Felipe Tejada-Castro$^{a}$}

\cortext[mycorrespondingauthor]{Corresponding author}
\ead{jcvasquez@pratechgroup.com}

\address[mymainaddress]{Pratech Group, Medell\'in, Colombia.}

\begin{abstract}

Customer service chatbots are conversational systems designed to provide information to customers about products/services offered by different companies. Particularly, intent recognition is one of the core components in the natural language understating capabilities of a chatbot system. Among the different intents that a chatbot is trained to recognize, there is a set of them that is universal to any customer service chatbot. Universal intents may include salutation, switch the conversation to a human agent, farewells, among others. A system to recognize those universal intents will be very helpful to optimize the training process of specific customer service chatbots. We propose the development of a universal intent recognition system, which is trained to recognize a selected group of 11 intents that are common in 28 different chatbots. The proposed system is trained considering state-of-the-art word-embedding models such as word2vec and BERT, and deep classifiers based on convolutional and recurrent neural networks. The proposed model is able to discriminate between those universal intents 
with a balanced accuracy up to 80.4\%. In addition, the proposed system is equally accurate to recognize intents expressed both in short and long text requests. At the same time, misclassification errors often occurs between intents with very similar semantic fields such as farewells and positive comments.   
The proposed system will be very helpful to optimize the training process of a customer service chatbot because some of the intents will be already available and detected by our system. At the same time, the proposed approach will be a suitable base model to train more specific chatbots by applying transfer learning strategies.

\end{abstract}


\end{frontmatter}

%\linenumbers

\section{Introduction}
\label{sec:introduction}

Customer service is necessary for companies from different sectors to attend their customer needs, and at the same to attract new ones~\cite{cui2017superagent}. Each customer service agent spends time solving different requests from customers about products or services offered by each company. Nevertheless, as the number of customers increases the waiting time increases as well, which results in poor customer satisfaction~\cite{ranoliya2017chatbot}. Traditional customer service has two main weaknesses: (1) agents usually receives repetitive requests asked by a variety of customers and (2) it is difficult to support services 24/7, especially for small companies~\cite{cui2017superagent}. With the aim of solving such issues, chatbots are able to handle multiple users simultaneously and operate all day and night long~\cite{ranoliya2017chatbot}.

Customer service chatbots are conversational software systems to automatically interact with a user/customer, providing accurate information about processes, services, and products offered by a specific company.  They are designed to emulate the communication capabilities of a human agent from customer service~\cite{nuruzzaman2018survey,jenkins2007analysis}. One of the core components of a suitable and accurate customer service chatbot is the natural language understanding (NLU) unit~\cite{pichl2020alquist}, which typically consists of different subsystems such as entity extraction, the dialogue management system, and intent recognition~\cite{lorenc2021benchmark}.  Particularly, intent recognition is a specific field of text classification, which is focused on finding the semantically closest intent to the request made to the chatbot by the user in each interaction~\cite{lorenc2021benchmark}.

Among the different intents that are considered to train a chatbot system, there is a set of them that is universal to any customer service chatbot. Universal intents may include salutation, switch the conversation to a human agent, change password in the company's platform, farewells, to request petitions and complaints, to give positive and negative comments, among others.  These types of intents are included in almost every customer service chatbot. A system to recognize a set of those universal intents will be very helpful to optimize the training process of a specific customer service chatbot because some of the intents used by the chatbot will be already available and detected by the universal intent recognizer. At the same time, a universal intent recognition system will be a suitable base model to train more specific chatbots by the use of transfer learning strategies. 

The purpose of this study is to propose the development of a universal intent recognition system, which is trained to recognize a set of intents that are usually considered in multiple customer service chatbot systems. The proposed system is trained considering state-of-the-art word-embedding models and deep learning based classifiers. To the best of our knowledge, this is one of the first approaches to recognize different intents that can be applied to multiple chatbot systems at the same time. 

\subsection{Related work}

Intent recognition systems have been accurately modeled by the use of recurrent neural networks (RNN), particularly using gated recurrent units (GRUs) and long short-term memory (LSTM) cells~\cite{chen2019bert,goo2018slot}. More recently, the use of Transformer models~\cite{vaswani2017attention} such as Bidirectional Encoder Representation from Transfomers (BERT)~\cite{devlin2018bert} have achieved state-of-the-art results for intent classification problems.

The authors in~\cite{goo2018slot} proposed a model to learn the relation between intents and entities by introducing a slot-gated mechanism into an attention layer combined with an LSTM network. The authors evaluated their proposed approach classifying 21 intents from the Airline Travel Information Systems (ATIS) dataset~\cite{tur2010left}, which has 4,478 utterances of people making flight reservations, and reported an accuracy of up to 94.1\%. 
The authors in~\cite{chen2019bert} explored the use of pre-trained BERT embeddings for joint intent classification and entity extraction. The authors reported an accuracy of up to 97.9\% recognizing 21 intents from the ATIS corpus. 
Another joint intent and entity classification was recently proposed in~\cite{lorenc2021joint}. The proposed model consisted on the computation of word-embeddings such as Glove, Fasttext, and BERT, which are classified using BiLSTM and time-distributed dense layers. The model was trained using a multi-task learning strategy to recognize at the same time intents and entities from the input's utterance. The authors reported an accuracy of up to 95.7\% classifying intents from the ATIS dataset.
In~\cite{zhang2021dialoguebert}, the authors proposed what they called a contextual dialogue encoder based on BERT (DialogueBERT) to address different classification problems that appear in dialogue systems such as intent recognition, entity extraction, and emotion classification. The proposed model consisted on different input embeddings, the same Transformer encoder from BERT, and a layer to encode information from full dialogue using a convolutional layer.  The authors reported an accuracy of up to 88.6\% for intent recognition, using a dataset with 185,200 requests labeled with 102 intents.  

Additional studies are focused on comparing the performance of cloud-based providers of NLU services~\cite{liu2021benchmarking,perez2021choosing}. For instance, the authors in~\cite{liu2021benchmarking} created a benchmark dataset for intent and entity recognition in human-robot interaction in at-home domains. The authors evaluated the performance of several cloud providers of NLU services such as IBM Watson assistant, Dialogflow, LUIS, and Rasa using the benchmark dataset. The results indicated that on intent classification Watson significantly outperforms the other platforms, achieving an accuracy up to 88.8\% recognizing 46 intents.

\section{Data}

\subsection{Spanish Customer Service Chatbots}

\begin{figure}[!ht]
        \setlength{\tabcolsep}{0pt} % Default value: 6pt
    \renewcommand{\arraystretch}{0} % Default value: 1
    \centering
    \begin{tabular}{cc}
        \textbf{a)} \\
        \includegraphics[width=\linewidth]{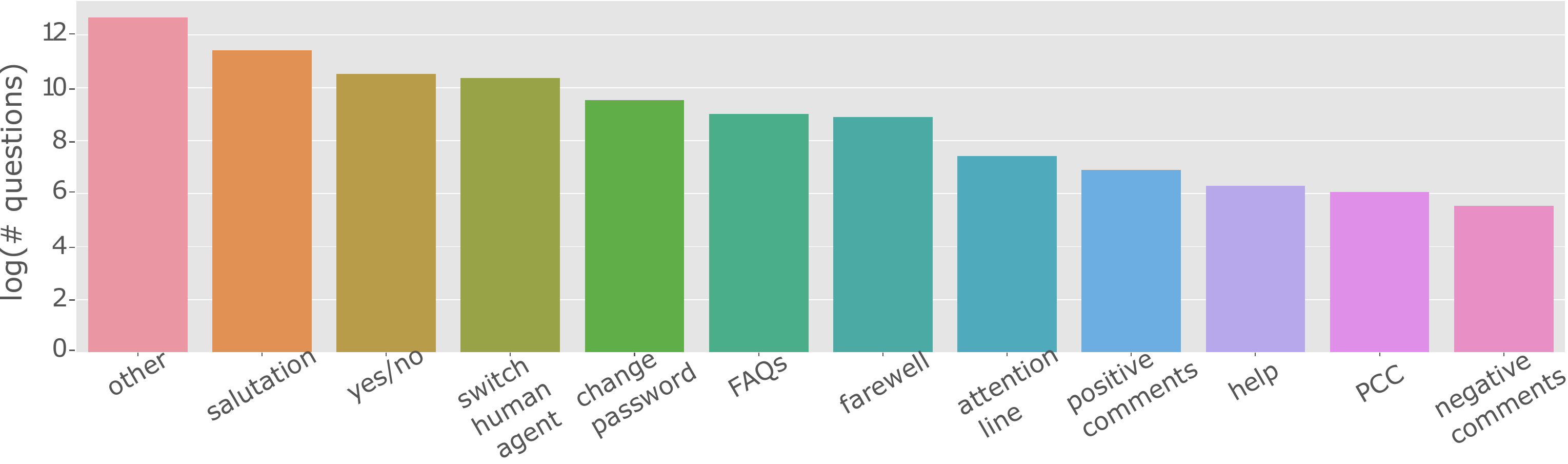}\\
        \textbf{b)} \\
        \includegraphics[width=\linewidth]{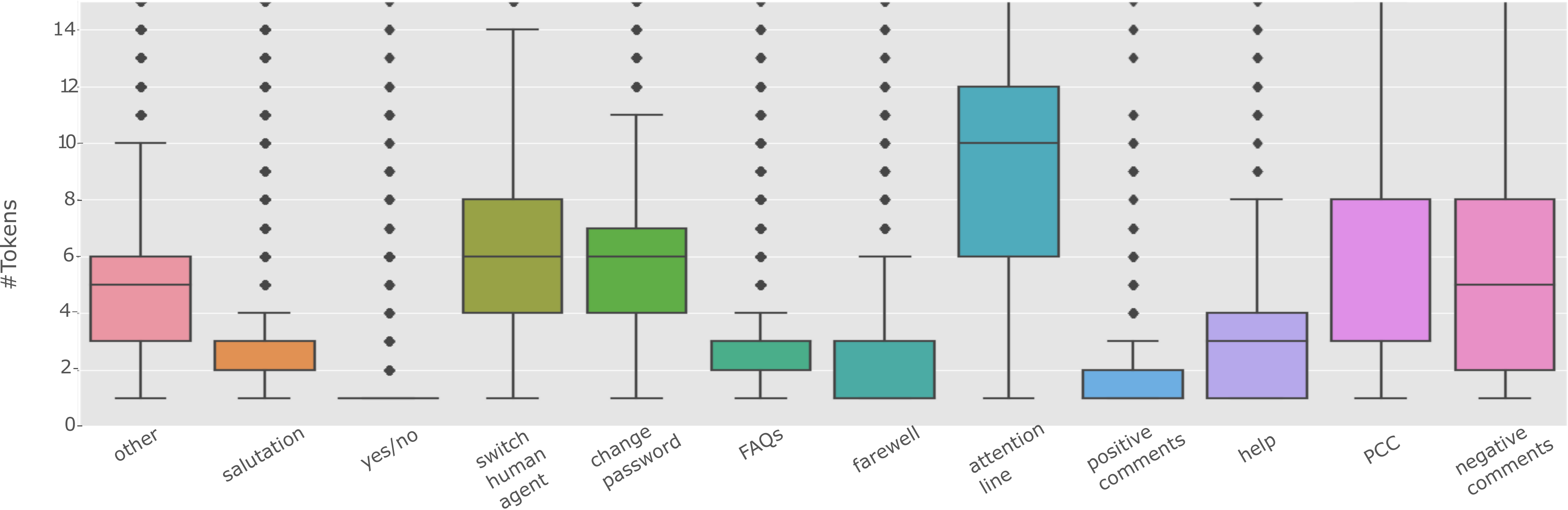}\\
        \textbf{c)} \\
        \includegraphics[width=\linewidth]{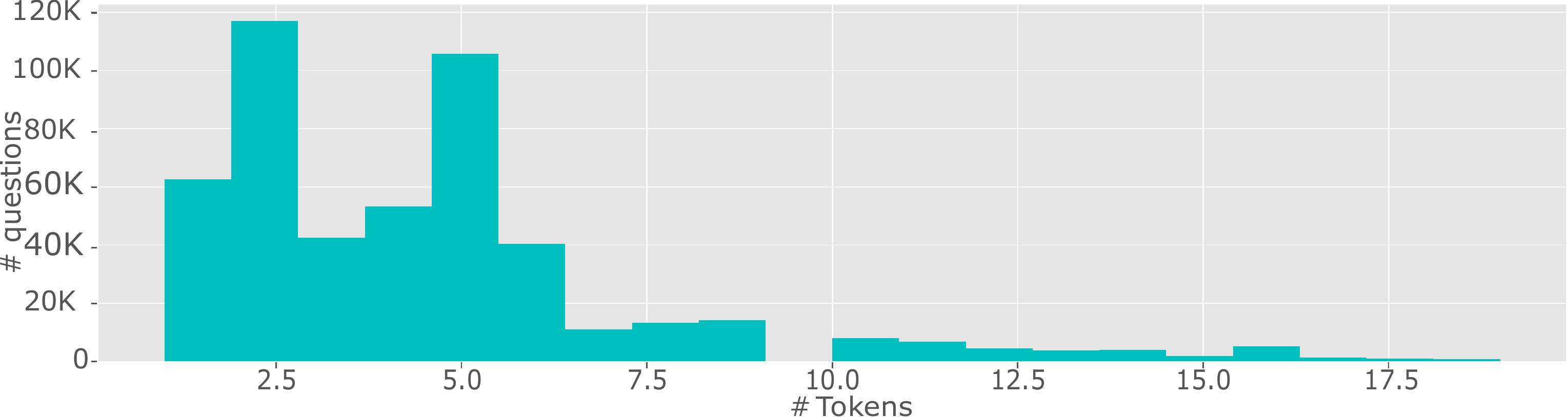}\\
    \end{tabular}
    
    \caption{Data distribution for the Spanish Customer Service Chatbots (SC$^2$) dataset. \textbf{FAQs}: frequently asked questions. \textbf{PCC}: Petitions, complaints, \& claims. \textbf{a)}. Distribution of samples per intent. \textbf{b)}. Number of tokens per intent. \textbf{c)} Histogram of the number of tokens in the requests to the chatbot systems.}
    \label{fig:data1}
\end{figure}

The Spanish Customer Service Chatbots (SC$^2$) dataset comprises data from 28 chatbot systems that are already in production and which are trained to provide customer service to different Colombian companies in several industries such as finance, insurance, health-care, retail, among others. We collected the requests made to each chatbot over a period of one month. We selected intents that appear at least in four chatbot systems and found a total of 11 intents that might be considered universal for customer service chatbots. The list of such universal intents is found in Table~\ref{tab:data1}. We carefully review the data and re-labeled the intents for each chatbot according to the 11 universal intents and leave the rest as \emph{others}. We collected a total of 506,823 requests made to 28 chatbots. The vocabulary for this corpus is formed with 34,310 unique words.
The distribution of samples across the different intents is shown as well in Table~\ref{tab:data1} and Figure~\ref{fig:data1}a). The number of requests representing universal intents across multiple chatbots is 192,446, which represent 38$\%$ of the total requests made to the chatbot systems (506,823). The data are randomly divided into train and test sets in an stratified fashion using 80$\%$ for training and the remaining 20$\%$ for test. Additionally, 10$\%$ of the train set is also randomly selected for the validation set, i.e., hyper-parameter optimization.

\begin{table}[!ht]
\centering
 \caption{Distribution of intents for the Spanish Customer Service Chatbot dataset. \textbf{FAQs}: frequently asked questions. \textbf{PCC}: Petitions, complaints, \& claims.}
 \resizebox{\linewidth}{!}{
\begin{tabular}{lrrrrrr}
\toprule
                      &             &             &            & \multicolumn{3}{c}{\# Requests} \\
\cline{5-7}
Intent                & \# words     & \# tokens   &\# chatbots & Train set & Test set & $\Sigma$ \\
\midrule
Others                & 4.0$\pm$4.6  & 5.6$\pm$7.0 & 28          & 251,501                & 62,876                & 314,377       \\
Salutation            & 1.9$\pm$1.6  & 3.6$\pm$3.5 & 26          & 72,104                 & 18,026                & 90,130              \\
Yes/No                & 1.7$\pm$0.7  & 1.1$\pm$0.9 & 10          & 29,778                 & 7,444                 & 37,222              \\
Switch to human agent & 4.5$\pm$3.0  & 6.5$\pm$4.8 & 20          & 25,626                 & 6,407                 & 32,033              \\
Change password       & 5.7$\pm$4.6  & 7.2$\pm$7.0 & 12          & 10,994                 & 2,749                 & 13,743              \\
FAQs                  & 2.7$\pm$1.4  & 3.3$\pm$2.6 & 4           & 6,538                  & 1,634                 & 8,172               \\
Farewell              & 2.0$\pm$2.4  & 2.4$\pm$4.0 & 14          & 5,803                  & 1,451                 & 7,254               \\
Attention line        & 7.7$\pm$7.3  & 10.9$\pm$10.7 & 7         & 1,348                  & 337                   & 1,685               \\
Positive comments     & 2.5$\pm$3.6  & 3.1$\pm$6.9 & 8           & 787                    & 197                   & 984                \\
Help                  & 3.7$\pm$3.6  & 4.2$\pm$6.0 & 4           & 433                    & 108                   & 541                \\
PCC                   & 4.0$\pm$5.2  & 6.7$\pm$6.8 & 4           & 341                    & 85                    & 426                \\
Negative comments     & 6.0$\pm$21.0  & 8.5$\pm$27.6 & 7         & 205                    & 51                    & 256                \\
\midrule
$\Sigma$              &             &               &           & 405,458                 & 101,365                & 506,823    \\
\bottomrule
\end{tabular}}
\label{tab:data1}
\end{table}

The distribution of the number of tokens for each intent of the corpus is shown in Figure~\ref{fig:data1}b). Note that there are intents with very short requests such as Yes/no, Farewell, Salutation, and FAQs, while others like Petitions, complaints, \& claims (PCC), negative comments, and asking for the attention line contain a larger number of tokens. This is expected having in mind that when customers are not satisfied with the service (PCC or negative comments), they almost always use more words to describe their bad experiences with the offered service.  Figure~\ref{fig:data1}c) shows the histogram of the number of tokens that are present in the corpus for each request. The average number of tokens present for each sentence ranges from 1 to 20, with an average of 5$\pm$5.5. All data are padded to the 95th percentile of the number of tokens (13) in order to be used as input for the considered deep learning architectures.

\subsection{Natural Language Data for Human-Robot Interaction in Home Domain}
\label{sec:dataeng}
\begin{figure}[!ht]
    \centering
    \begin{tabular}{c}
        \textbf{a)} \\
        \includegraphics[width=\linewidth]{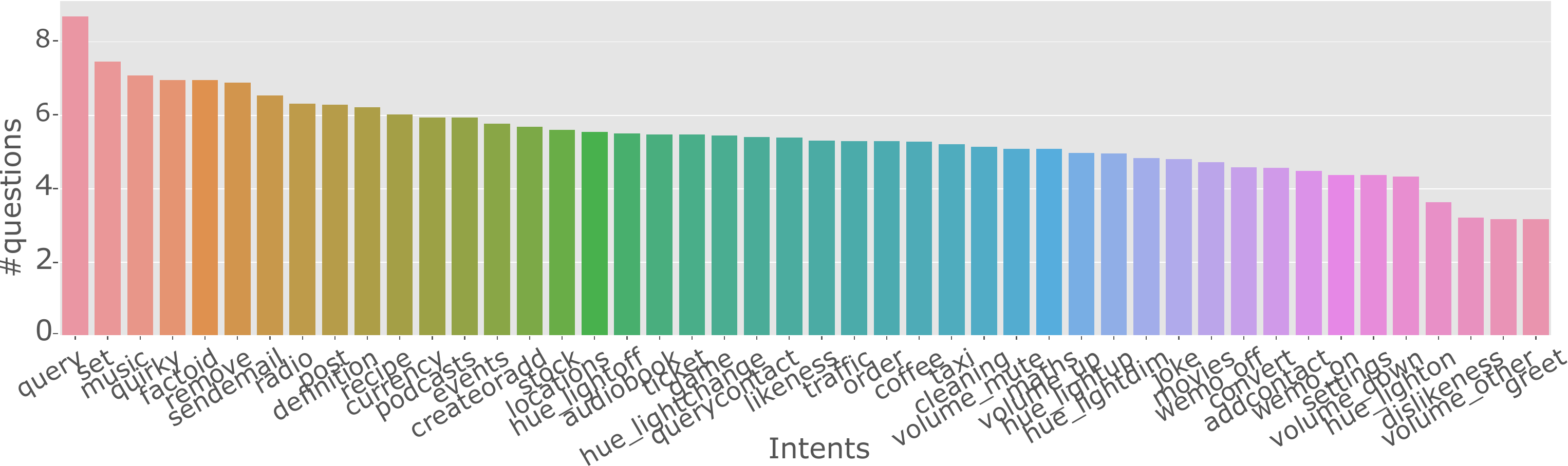} \\
        \textbf{b)}\\
        \includegraphics[width=\linewidth]{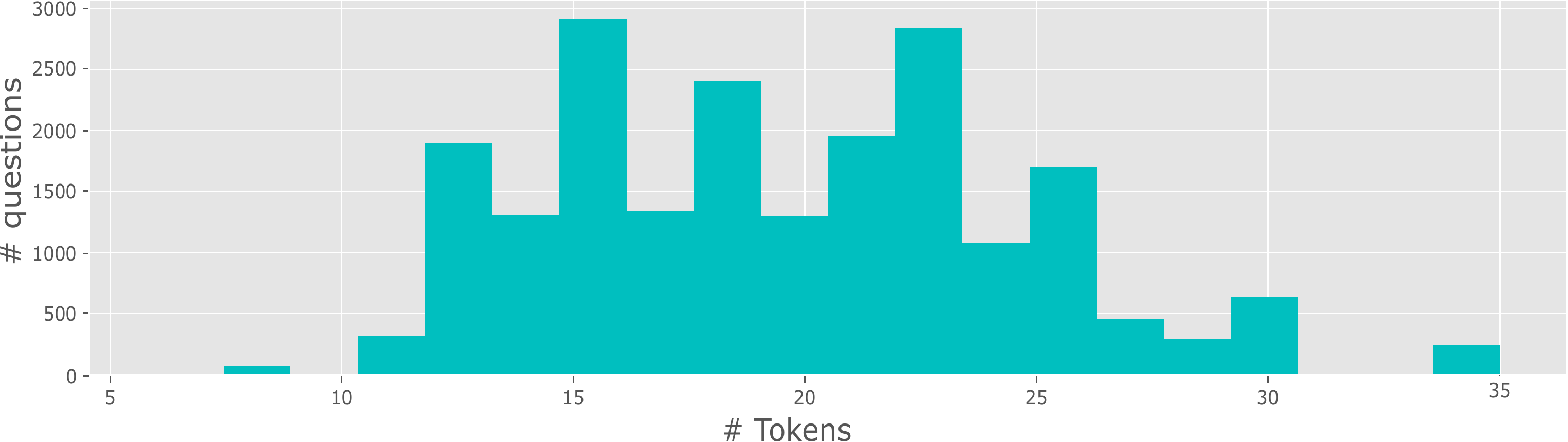}  \\
    \end{tabular}
    
    \caption{Data distribution for the Natural Language Data for Human-Robot Interaction in Home Domain dataset. \textbf{a)}. Distribution of samples per intent. \textbf{b)}. Number of tokens in the sentences. }
    \label{fig:data2}
\end{figure}

The considered models to train the intent recognition system are validated in a second corpora.  We considered the Natural Language Data for Human-Robot Interaction in Home Domain~\cite{liu2021benchmarking}, and which is publicly available\footnote{\url{https://github.com/xliuhw/NLU-Evaluation-Data}}. The dataset comprises 25,716 requests made to a personal assistant system, and which are labeled for 46 different intents, after removing empty requests. Figure~\ref{fig:data2}a) shows the distribution of the samples per intent. The average number of tokens present for each sentence ranges from 4 to 35 (see Figure~\ref{fig:data2}b)), with an average of 19.6$\pm$5. All data are padded as well to the 95th percentile of the number of tokens (28) in order to be used as input for the considered deep learning architectures. The original train test split is used for the experiments addressed here. In addition, 10\% of the training set (randomly chosen) is used as the validation set for hyper-parameter optimization.

\section{Methods}

\subsection{Word embeddings}

Given a set of words $W = \left[ w_1, w_2, w_3, ..., w_V \right]$, where $V$ is the vocabulary size, a word-embedding is a vector representation where each point corresponds to a word in the vocabulary and the similarity between each pair of points is well defined~\cite{yan2009text}.
Word embeddings are real-valued representations of words produced by distributional semantic models~\cite{bakarov2018survey}. 
There are different methods and architectures to generate word embeddings and these can be either context-independent or context-dependent. This study considers both context-independent and context-dependent word embedding models to evaluate our proposed system. Details of each representations are presented in the following subsections.

\subsubsection{Context-independent: Word2vec}

Word2vec is one of the most used embedding methods in the literature~\cite{li2019scaling}. This model is built on top of a large  corpus in order to produce a vector space, typically of several hundred dimensions. Word vectors are positioned in the vector space in such a way that words sharing semantic similarities in the corpus are geometrically close to each other~\cite{mikolov2013distributed}. 
The word2vec model is designed to find a unique vector to represent each word in the corpus. For this reason it is known as a context-independent embedding, i.e.,  the representation of a word is the same regardless of its context. 
These types of representations have gained popularity not only because each word-embedding keeps the semantic properties of the word, but also because the model is trained in an unsupervised fashion, i.e., the texts that are analyzed do not require prior labeling. Hence, it is possible to train models with a large amount of data based on freely accessible text resources without spending money on expensive hand-labeled databases.

The algorithm to train a word2vec model uses a neural network to learn word relations. The words in the corpus are transformed into a one-hot representation to feed the network. There are two common strategies in the literature to built the structure of the neural network used to learn the word2vec representations: (1) Skip Gram and (2) Continuous Bag Of Words (CBOW). We consider the CBOW architecture for the models introduced here because we have observed that such strategy produces more accurate representations when large corpora are available~\cite{alwehaibi2018comparison,bhoir2017comparative}. CBOW takes the 
context of each word (one-hot encoded) as input and the network aims to predict the word corresponding to the context. The number of context words is previously defined, and typically ranges between 3 and 7 words~\cite{caselles2018word2vec}. 

We considered separate word2vec models for each corpus used in our experiments. On one hand, regarding the SC$^2$ dataset, we trained two different word2vec models. The first one was trained using the Spanish WikiCorpus, which contains 
120 million words~\cite{reese2010word} and using the CBOW strategy with 7 context words, and using 100 dimensions to represent each word. The aim of this model is to evaluate the performance of a general model that can be adapted to other NLU tasks not directly related to the target vocabulary that is available in chatbot interactions. This model will be referenced as \emph{w2v-wiki}. The second word2vec model is directly trained with the 34,310 words that appear in the chatbot requests from the SC$^2$ data in order to test whether the specific vocabulary  used in service chatbots is better to learn most accurate semantic relationships among the words in the corpus. This model is also trained to obtain a 100-dimensional feature vector per word, and using a context of 4 words because that is the average number of words in the requests made to the chatbots. This model will be referred as \emph{w2v-sc2}. On the other hand, for the English corpus, we considered as well two word2vec models: (1) the pre-trained word2vec-google-news-300 model, which is available online\footnote{\url{https://code.google.com/archive/p/word2vec/}}, and which consists of 300-dimensional feature vectors trained with the Google News corpus (about 100 billion words) with a context of 5 words~\cite{mikolov2013distributed,mikolov2013efficient}. This model will be referred as \emph{w2v-google}. (2) A specifically created word2vec model trained with the 311 unique words from the chatbot data described in Section~\ref{sec:dataeng}, and using 8 words of context. This models will be referred as \emph{w2v-bot}.

\subsubsection{Context-dependent: BERT}

In context-dependent models, the embedding obtained for a certain word is not unique, it will depend on the context of the word in the sentence. Therefore the same word may have different embedding representations depending on the context the word appear in the request to the chatbot.  Recent developments of context-dependent embeddings~\cite{devlin2018bert,peters2018deep} show that systems based on such representations achieve good results in different text classification tasks~\cite{miaschi2020contextual}, including fake news detection~\cite{martinez2021fake}, detection of user satisfaction in chatbot systems and call centers~\cite{escobar2021,perez2021user}, document classification~\cite{adhikari2019docbert}, and  health-care applications~\cite{perez2021user}.

\begin{figure}[!ht]
\centering
 \includegraphics[width=0.4\textwidth]{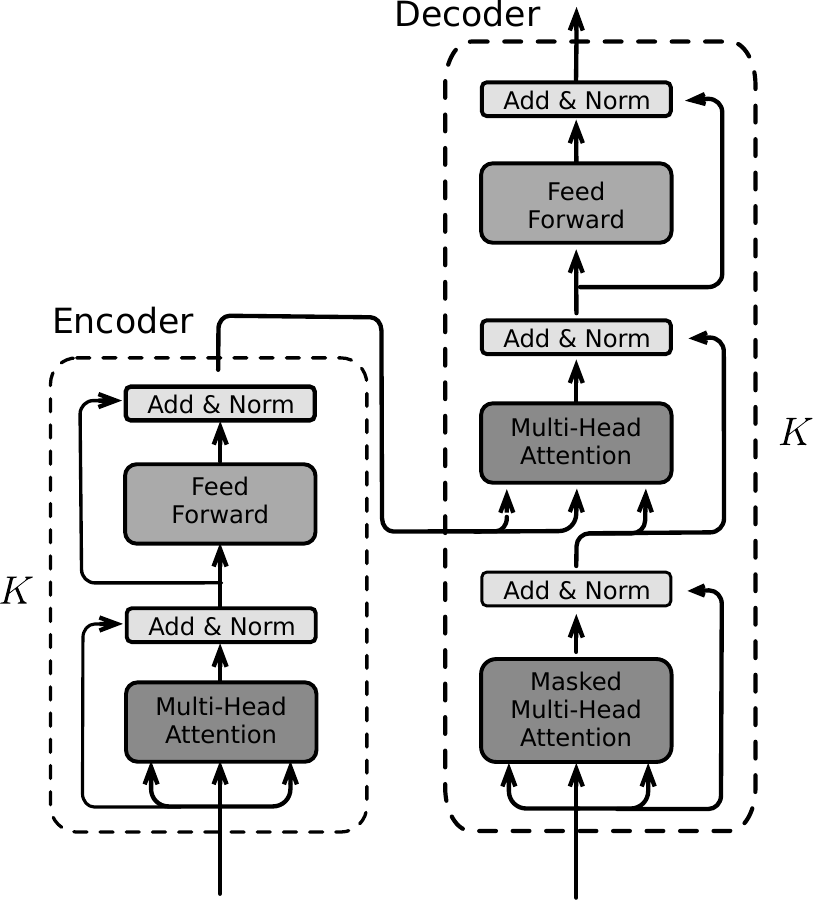}
\caption{Topology of the Transformer architecture. 
$K=12$ is the number of layers in the encoder and decoder. 
Figure is adapted from~\cite{vaswani2017attention}.}
\label{fig:Transformermethodology}       % Give a unique label
\end{figure}

BERT~\cite{devlin2018bert} is one the most popular context-dependent embeddings. It is based on a Transformer architecture~\cite{vaswani2017attention}, originally created for machine translation. The Transformer includes two separate mechanisms: an encoder to receive the input's text and a decoder to produce  a prediction for the task. The encoder is formed with a stack of layers that include self-attention and feed-forward connections. Decoders include all the elements present in the encoder with an additional encoder-decoder attention layer between the self-attention and the feed-forward layers~\cite{vaswani2017attention}. Figure~\ref{fig:Transformermethodology} shows the encoder and decoder of the Transformer architecture. Only the encoder part is used to compute the BERT embeddings. The most important aspect in the model is the multi-head attention mechanism. This mechanism consists of several attention layers running in parallel in order to learn contextual relations among words (or sub-words) in a text. The attention function is defined according to Equation~\ref{eq_attention1}. $\mathbf{Q}$, $\mathbf{K}$, and $\mathbf{V}$ are known as the query, key, and value matrices, respectively. 

\begin{equation}
    \mathrm{Attention}(\mathbf{Q}, \mathbf{K}, \mathbf{V}) = \mathrm{softmax}\left(\frac{\mathbf{Q}\mathbf{K}^{\intercal}}{\sqrt{d_k}}\right)\textbf{V}
\label{eq_attention1}
\end{equation}

These matrices are built from the word-embeddings of sequences extracted from the text. In BERT, $\mathbf{Q}$, $\mathbf{K}$, and $\mathbf{V}$ are the same matrix. This is why the mechanism is known as self-attention in BERT embeddings. The dot product between $\mathbf{Q}$ and $\mathbf{K}^{\intercal}$ has information about the relationships among elements of $\mathbf{Q}$ and elements of $\mathbf{K}$. Then, the product is scaled by $\sqrt{d_k}$, where $d_k$ is the dimension of the word-embedding. Finally, a softmax function is applied to obtain posterior probabilities, which are multiplied with the $\textbf{V}$ matrix. The posterior probabilities obtained from the Softmax function give importance to each component of  $\textbf{V}$, obtaining a word embedding for each word as a linear combination of the context words in the sentence.

We considered a pretrained BERT model for all our experiments. We used the BERT-Base Multilingual cased pre-trained model, which was trained with the Multi-Genre Natural Language Inference (MultiNLI) corpus.
The architecture of the BERT-Base model consists of 12 self-attention layers each one with 768 hidden units. The last layer (768 units) is taken as the word-embedding representation. The pre-trained model is available to be downloaded in Tensorflow hub\footnote{\url{https://tfhub.dev/tensorflow/bert_multi_cased_L-12_H-768_A-12/4}}.

\subsection{Classification Models}

\begin{figure*}[!ht]
\centering
\begin{tabular}{c}
 \textbf{a)}\\
 \includegraphics[width=0.7\textwidth]{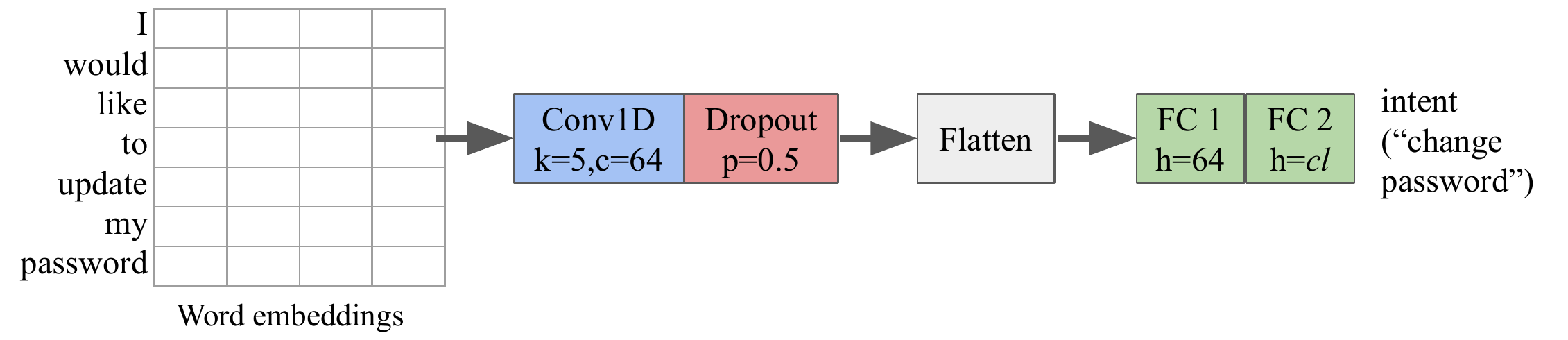}\\
  \textbf{b)}\\
 \includegraphics[width=0.7\textwidth]{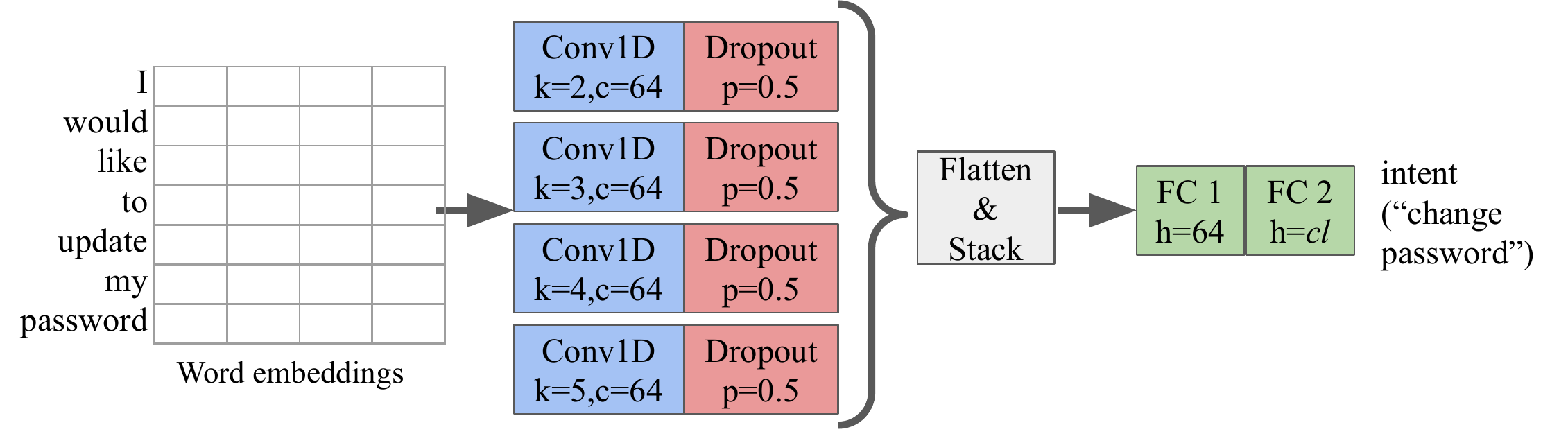}\\  
  \textbf{c)}\\
 \includegraphics[width=0.7\textwidth]{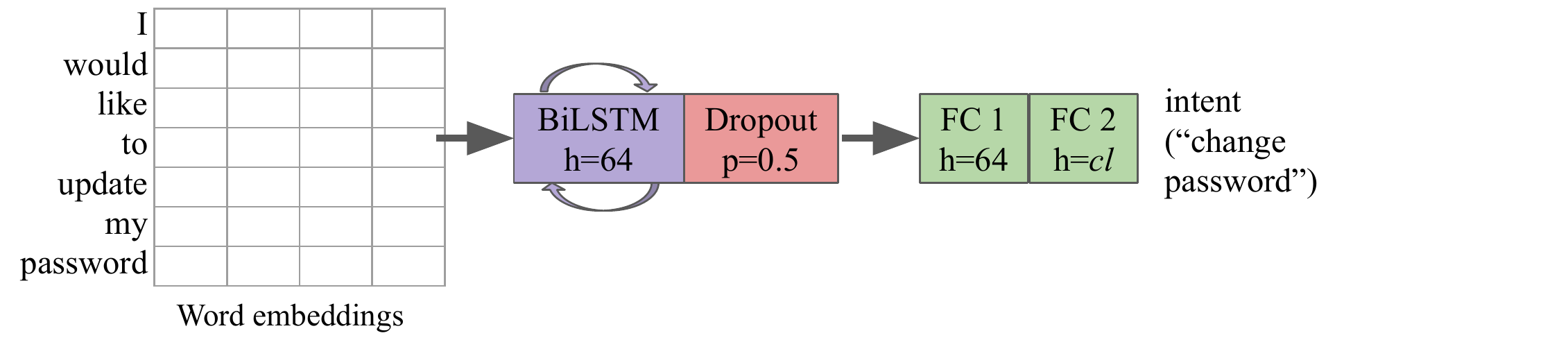}\\
   \textbf{d)}\\
 \includegraphics[width=0.7\textwidth]{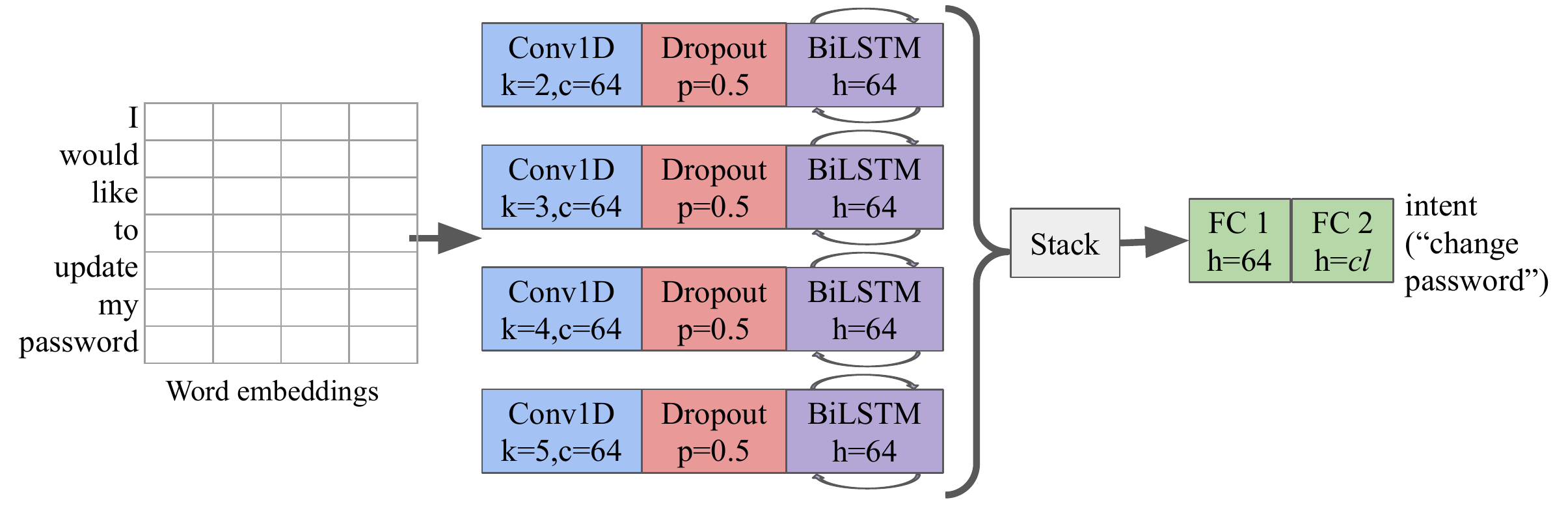}
\end{tabular}
\caption{Architecture of the deep learning models for intent recognition: \textbf{a)} 1D-CNN. \textbf{b)} Parallel 1D-CNN \textbf{c)} BiLSTM. \textbf{d)} Parallel 1D-CNN followed by BiLSTM layers. \textbf{k}: kernel size for the convolution layers, \textbf{c}: number of output filters for the convolution layers, \textbf{FC}: fully connected layer, \textbf{p}: dropout probability, \textbf{h}: number of hidden neurons, \textbf{\textit{cl}}: number of classes.}
\label{fig:models}       % Give a unique label
\end{figure*}

The set of features extracted using the considered word embeddings are used to train four different deep learning-based classifiers combining convolution and recurrent layers.  The four considered architectures are shown in Figure~\ref{fig:models}. 

The first model (see Figure~\ref{fig:models}a)) comprises a simple 1D-CNN followed by 2 fully connected layers. The aim of the 1D convolution layer is to extract semantic features for the complete request via the convolution operation between the word embedding matrix and a set of learned filters (c=64) with a fixed temporal resolution (k=5 tokens). A ReLU activation function is used in the convolution and the first fully connected layers. The output layer includes a Softmax activation function to get posterior probabilities of the recognized intents. 

The second model in Figure~\ref{fig:models}b) is a parallel CNN where convolution layers with different temporal resolutions (filter's order) are considered. Different filter orders correspond to different number of $n$ 
in $n$-grams. Filters with  $2$nd, $3$rd, $4$th, and $5$th order are designed to map different semantic relationships including 
\textit{bi}-gram, \textit{tri}-gram, \textit{four}-gram, and \textit{five}-gram respectively. 
Parallel convolution layers have been successfully applied in similar text classification tasks related to effectiveness analysis in chatbot systems~\cite{escobar2021}, and user profiling~\cite{escobar2021gender}.
The outputs of the four convolution layers are flattened and stacked together to feed 2 fully connected layers to make the classification. The same activation functions that are considered in simple 1D-CNN model are also considered for this architecture. 

The third considered model, which is shown in Figure~\ref{fig:models}c) comprises an RNN with BiLSTM cells with the aim to model the temporal context of the request to the chatbot system.
LSTMs have a \emph{causal} structure, i.e., the output  at the present time step only contains information from the past.  However, many applications require information from the future~\cite{otter2020survey}. BiLSTMs cells are created to address such a requirement by combing a layer that processes the input sequence forward through time with an additional layer that moves backwards the input sequence. Word embedding matrices from the chatbot's requests are used as the input for the BiLSTM layer, which is trained in a many-to-one scenario, i.e., only the output at the last time step of the layer is considered because it carries information about the whole text sequence. The output of the BiLSTM layer feeds two fully connected layers to made the final decision, using the same activation functions than in the previous two architectures.  

Finally, the fourth model shown in Figure~\ref{fig:models}d) comprises the combination of parallel convolutions with BiLSTM layers to model the temporal context of the text sequence. The output of the BiLSTM layer is passed as well to the two fully connected layers with the same characteristics of the previous models in order to make the final classification.

\subsubsection{Training}

The classification models are implemented in Tensorflow 2.0, and are trained with a sparse categorical cross-entropy loss function using an Adam optimizer with a learning rate of 0.0001 for 50 epochs.  Class weights were used to compute the loss function in order to compensate the unbalanced classes in both corpora. The class weight $w_i$ for the $i$-th class is computed using Equation~\ref{eq:weight}, where $N$ is the size of the training data, $C$ is the number of classes, i.e., intents, and $n_i$ is the number of samples of the $i$-th class in the training set.

\begin{equation}
    w_i= \frac{N}{C*n_i}
    \label{eq:weight}
\end{equation}

\section{Results}

The results obtained recognizing intents in the SC$^2$ dataset, which are universal for multiple chatbot systems are shown in Table~\ref{tab:results_spanish} using the different word embeddings and classification models. The highest unweighted average recall (UAR)  (80.4\%) is obtained using the specific word2vec model trained with the chatbot data classified with the paralle CNN model. Note that relatively similar accuracies are observed independent on the classification model. The results are more dependent on the considered word embedding model than on the classification network. 
The highest accuracies are obtained using the specific word2vec model trained with the chatbot data, followed by the general word2vec model trained on the wikipedia corpus. Finally the lowest results are obtained with the BERT embeddings.

\begin{table}[!ht]
\centering
 \caption{Results recognizing intents in the SC$^2$ dataset for the different word-embeddings and classification models. \textbf{ACC}: accuracy, \textbf{UAR}: unweighted average recall. Model with the highest UAR is highlighted in bold.}
 \resizebox{\linewidth}{!}{
\begin{tabular}{lllccc}
\toprule
\textbf{word embedding} & \textbf{Classification model} & \textbf{\# params} & \textbf{ACC} & \textbf{UAR} & \textbf{F-score} \\ \midrule
w2v-wiki                & CNN                           & 49,202              & 87.2         & 77.3         & 54.6             \\
w2v-wiki                & parCNN                       & 180,812             & 87.7         & 77.2         & 54.8             \\
w2v-wiki                & BiLSTM                        & 251,679             & 89.7         & 79.0         & 56.8             \\
w2v-wiki                & parCNN+BiLSTM                 & 387,595             & 87.3         & 79.2         & 55.5             \\
w2v-sc2                 & CNN                           & 49,202              & 89.1         & 80.2         & 55.8             \\
\textbf{w2v-sc2}        & \textbf{parCNN}              & \textbf{180,812}    & \textbf{89.3} & \textbf{80.4} & \textbf{55.9}  \\
w2v-sc2                 & BiLSTM                        & 251,679             & 89.0         & 80.0         & 55.9             \\
w2v-sc2                & parCNN+BiLSTM                 & 387,595             & 89.1         & 80.7         & 55.8             \\
BERT                    & CNN                           & 283,532             & 84.0         & 75.0         & 51.5             \\
BERT                    & parCNN                       & 861,260             & 83.2         & 77.5         & 51.4             \\
BERT                    & BiLSTM                        & 937,966             & 85.0         & 78.0         & 53.0             \\ 
BERT                    & parCNN+BiLSTM                 &  986,188            &  81.5        & 76.4         & 50.4             \\ 

\bottomrule
\end{tabular}}
\label{tab:results_spanish}
\end{table}

Details of the classification of each intent in the SC$^2$ corpus are shown in Table~\ref{tab:results_spanish2} for the most accurate model i.e., word2vec trained with the chatbot data classified using a parallel CNN. Intents are sorted according to the number of samples in the corpus (see Table~\ref{tab:data1} and Figure~\ref{fig:data1}a)). The most accurate intents to be recognized are \emph{Yes/No, Switch to human agent, Change password, Salutation, and Others}. Relatively high accuracies (higher than 80\%) are also observed for \emph{FAQs, Attention line, Positive comments, and PCC}. Unfortunately, some of the under sampled intents such as \emph{Help} and \emph{Negative comments} are not very accurate, particularly because the low number of samples available for such classes.

\begin{table}[!ht]
\centering
 \caption{Details of the most accurate model to recognize intents that are universal to multiple service chatbots in the SC$^2$ dataset.}
 \resizebox{\linewidth}{!}{\begin{tabular}{lccc}
\toprule
\textbf{Intent}       & \textbf{Precision (\%)} & \textbf{Recall (\%)} & \textbf{F-score (\%)} \\ \midrule
Others                & 99.1      & 87.9   & 93.2    \\
Salutation            & 97.4      & 91.3   & 94.3    \\
Yes/No               & 92.8      & 98.5   & 95.6    \\
Switch to human agent  & 96.2      & 94.2   & 95.2    \\
Change password       & 67.8      & 93.8   & 78.7    \\
FAQs                  & 51.3      & 86.1   & 64.3    \\
Farewell              & 61.1      & 59.5   & 60.3    \\
Attention line        & 12.1      & 84.0   & 21.2    \\
Positive comments     & 5.3       & 80.2   & 9.9     \\
Help                  & 12.3      & 42.5   & 19.1    \\
PCC                   & 12.1      & 82.3   & 21.2    \\
Negative comments     & 10.4      & 64.7   & 17.9    \\
 \bottomrule
\end{tabular}}
\label{tab:results_spanish2}
\end{table}

\begin{figure}[!ht]
    \setlength{\tabcolsep}{0pt} % Default value: 6pt
    \renewcommand{\arraystretch}{0} % Default value: 1
    \centering
    \begin{tabular}{cc}
    \textbf{a)} & \textbf{b)}\\
    \includegraphics[width=0.5\linewidth]{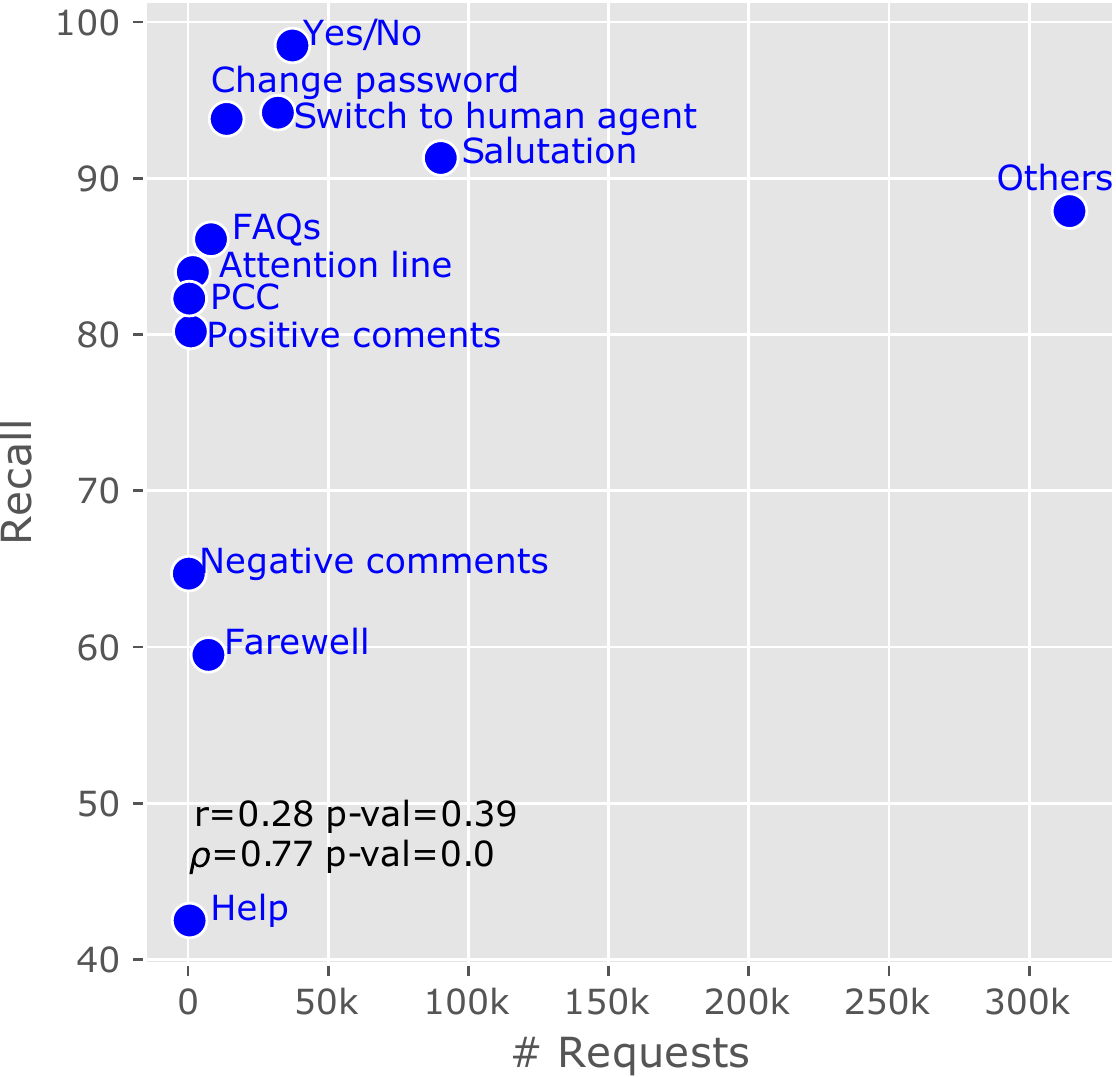}     &
    \includegraphics[width=0.5\linewidth]{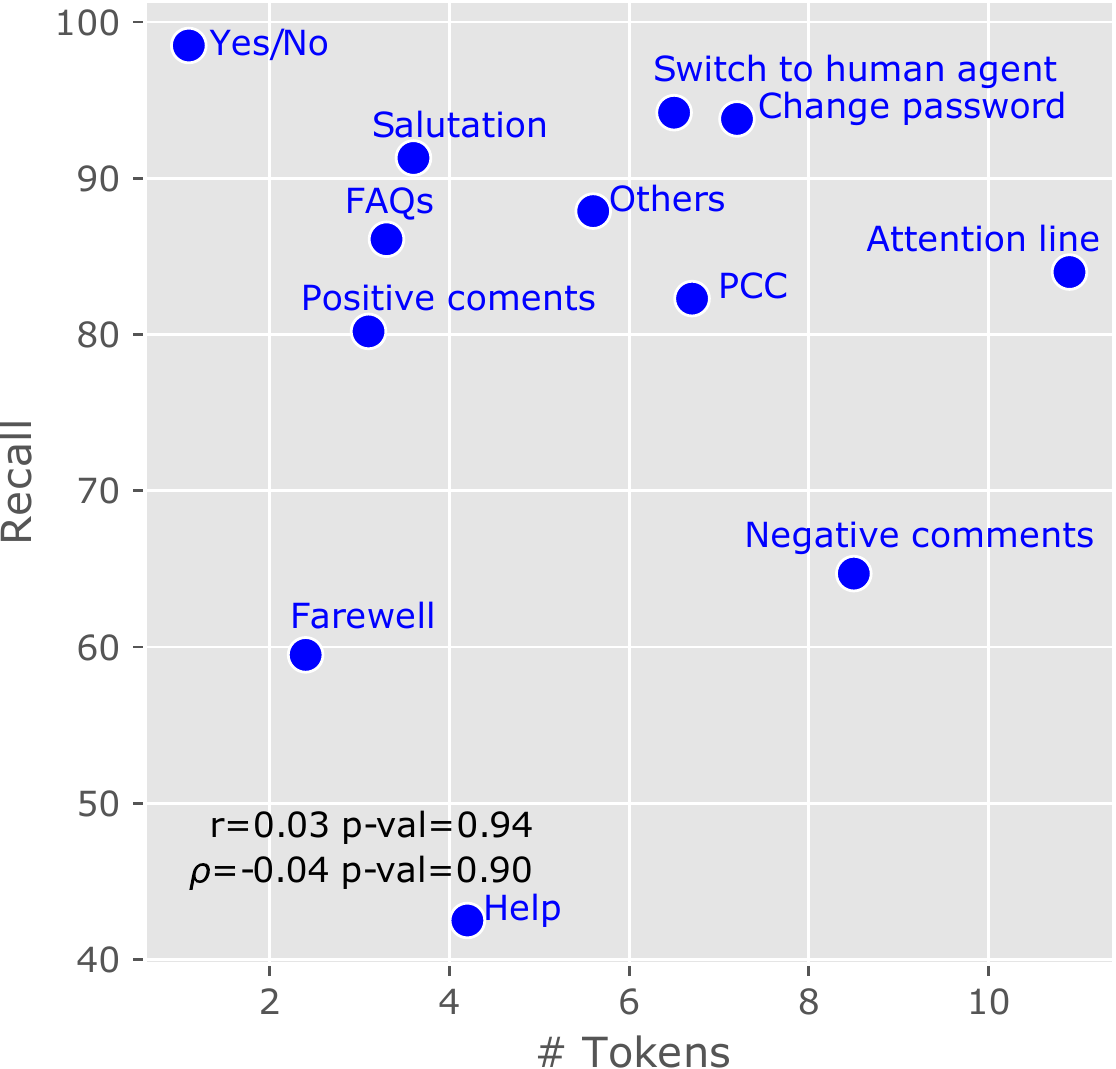}
    \end{tabular}
    \caption{Correlation between recall obtained for each intent in the in the SC$^2$ dataset and: \textbf{a)} number of requests for each intent. \textbf{b)} numbers of tokens for each intent.}
    \label{fig:analysis_spanish}
\end{figure}

The recall obtained for each intent is compared against the number of request that are available for each intent (see Figure~\ref{fig:analysis_spanish}a)) and the average number of tokens used to request each intent (see Figure~\ref{fig:analysis_spanish}b)). The Pearson's ($r$) and Spearman's ($\rho$) correlation coefficients are computed to evaluate the comparisons. On one hand, there is a positive  and significant Spearman's correlation ($\rho=0.77, \, p-val<0.0005$) between the number of requests and the recall obtained for each intent. The fact that the Pearson's correlation is much lower and non-significant ($\rho=0.28, \,p-val=0.39$) indicates that the existing correlation between the number of samples in the corpus and the recall is not linear. In order to improve the accuracy of the considered models, the number of samples of the under-represented classes should be highly increased. On the other hand, there is no correlation between the length of the requests used by the user, represented in the number of tokens, and the accuracy of the model to recognize each intent. The considered models are equally accurate to recognize both short and long requests made to the chatbot system.

\begin{figure}[!ht]
\centering
 \includegraphics[width=\linewidth]{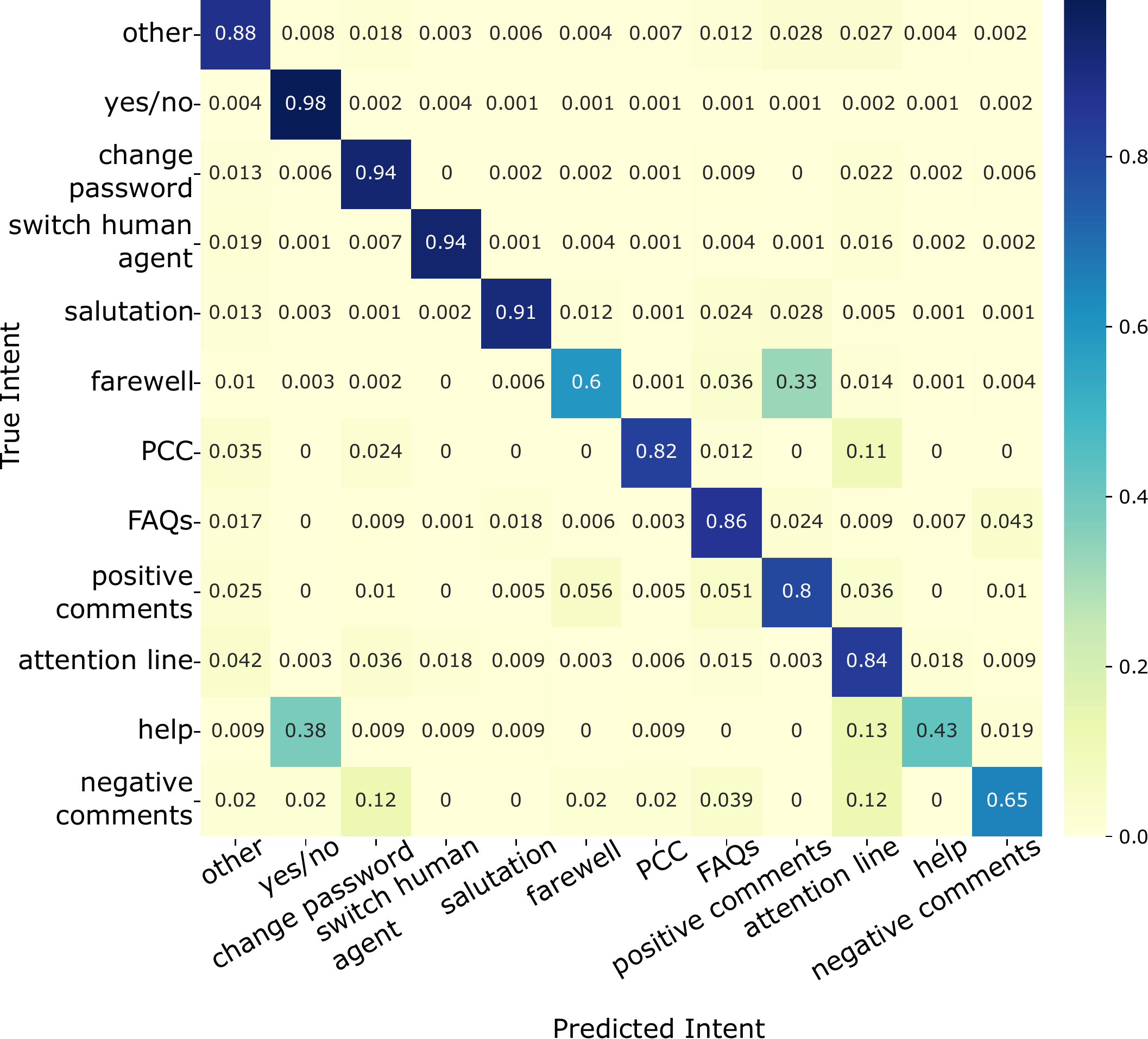}
\caption{Confusion matrix for the most accurate intent recognition system in the SC$^2$ dataset.}
\label{fig:cm_spanish}       % Give a unique label
\end{figure}

The confusion matrix for the most accurate model is shown in Figure~\ref{fig:cm_spanish}. These results confirm the high accuracies observed for most of the intents, except for those under sampled classes such as \emph{Help} and \emph{Negative comments}, which are not very accurate. Particularly, note that \emph{Farewells} are mainly misclassified with \emph{Positive comments}. This is because both intents share similar semantic fields related to acknowledge for the provided service at the end of the conversation. Such type of farewells are detected as positive comments by the system.

The results obtained for the English corpus are shown in Table~\ref{tab:results_english}. The model with the highest UAR corresponds to the pretrained word2vec from google-news classified with the CNN (UAR=82.5\%). In general, higher accuracies are observed for the models based on word2vec embeddings, similar to the previous case, excepting for the model based on BERT embeddings classified with the parallel CNN (UAR=82.0\%).

\begin{table}[!ht]
\centering
 \caption{Results recognizing intents in the Natural Language Data for Human-Robot Interaction for the different word-embeddings and classification models. \textbf{ACC}: accuracy, \textbf{UAR}: unweighted average recall. Model with the highest UAR is highlighted in bold.}
 \resizebox{\linewidth}{!}{
\begin{tabular}{llllll}
\toprule
\textbf{word embedding} & \textbf{Classification model} & \textbf{\# params} & \textbf{ACC} & \textbf{UAR} & \textbf{F-score} \\ \midrule
\textbf{w2v-google} & \textbf{CNN}    & \textbf{181,038}  & \textbf{87.2} & \textbf{82.5} & \textbf{76.4}    \\
w2v-google     & parCNN              & 624,336    & 87.4 & 81.3 & 74.5  \\
w2v-google     & BiLSTM               & 458,734    & 83.2 & 80.4 & 71.5 \\
w2v-google     & parCNN+BiLSTM        & 569,070    & 86.7 & 79.7 & 72.9 \\
w2v-bot        & CNN                  & 181,038    & 86.5 & 81.4 & 73.2 \\
w2v-bot        & parCNN              & 624,336    & 87.0 & 81.6 & 74.2  \\
w2v-bot        & BiLSTM               & 458,734    & 85.6 & 82.3 & 73.5 \\
w2v-bot         & parCNN+BiLSTM       & 569,070    & 86.0 & 80.8 & 73.4 \\
BERT           & CNN                  & 347,182    & 79.5 & 62.3 & 54.2 \\
BERT           & parCNN              & 1,109,230   & 88.0 & 82.0 & 76.0 \\
BERT           & BiLSTM               & 937,966    & 76.5 & 61.4 & 62.3 \\
BERT           & parCNN+BiLSTM        & 988,398    & 76.4 & 61.0 & 61.1 \\
\bottomrule
\end{tabular}}
\label{tab:results_english}
\end{table}

Details of the best resulting model to classify each intent in the corpus are shown in Table~\ref{tab:results_english2}. The intents are sorted according to the number of samples in the dataset. 18 of the 46 intents are very accurate to recognize (recall$>95\%$): \emph{post, cleaning, set, joke, game, radio, querycontact, currency, hue$\_$lightup, audiobook, traffic, order, coffee, createoradd, recipe, convert,} and \emph{taxi}. Conversely, intents such as \emph{volume$\_$down, hue$\_$lighton,} and \emph{volume$\_$other} are not recognized at all. The results also suggest that there could be a threshold in the  number of request for the intent to be accurately recognized.

\begin{table}[!ht]
\centering
 \caption{Details of the most accurate model to recognize intents in the Natural Language Data for Human-Robot Interaction}
 \resizebox{\linewidth}{!}{
\begin{tabular}{lccc|lccc}
\toprule
\textbf{Intent}       & \textbf{Precision} & \textbf{Recall} & \textbf{F-score} & \textbf{Intent}           & \textbf{Precision} & \textbf{Recall} & \textbf{F-score} \\ \midrule
query        & 94.4      & 82.9   & 88.2    & likeness         & 47.3      & 63.4   & 54.2    \\
set          & 98.2      & 95.4   & 96.8    & traffic          & 97.5      & 97.5   & 97.5    \\
music        & 98.1      & 87.9   & 92.7    & order            & 60.9      & 97.5   & 75.0    \\
quirky       & 68.7      & 69.7   & 69.2    & coffee           & 100.0     & 97.5   & 98.7    \\
factoid      & 86.4      & 81.4   & 83.8    & taxi             & 97.4      & 100.0  & 98.7    \\
remove       & 98.4      & 91.4   & 94.7    & cleaning         & 100.0     & 100.0  & 100.0   \\
sendemail    & 98.5      & 92.8   & 95.6    & volume\_mute     & 93.8      & 90.9   & 92.3    \\
radio        & 96.5      & 99.1   & 97.8    & maths            & 97.0      & 100.0  & 98.5   \\
post         & 99.1      & 99.1   & 99.1    & volume\_up       & 46.5      & 69.0   & 55.6    \\
definition   & 88.6      & 100.0  & 94.0    & hue\_lightup     & 84.8      & 100.0  & 91.8    \\
recipe       & 97.6      & 96.4   & 97.0    & hue\_lightdim    & 87.0      & 80.0   & 83.3    \\
currency     & 94.9      & 98.7   & 96.8    & joke             & 100.0     & 95.8   & 97.9    \\
podcasts     & 96.2      & 100.0  & 98.1    & movies           & 24.4      & 45.5   & 31.7    \\
events       & 73.5      & 39.1   & 51.0    & wemo\_off        & 77.8      & 35.0   & 48.3    \\
createoradd  & 96.7      & 100.0  & 98.3    & convert          & 25.7      & 100.0  & 40.9    \\
stock        & 100.0     & 87.0   & 93.1    & addcontact       & 73.9      & 94.4   & 82.9     \\
locations    & 95.6      & 84.3   & 89.6    & wemo\_on         & 51.9      & 87.5   & 65.1    \\
hue\_lightoff & 91.8     & 91.8   & 91.8    & settings         & 23.4      & 93.8   & 37.5   \\
audiobook    & 94.0      & 97.9   & 95.9    & volume\_down     & 0.0       & 0.0    & 0.0     \\
ticket       & 68.8      & 91.7   & 78.6    & hue\_lighton     & 0.0       & 0.0    & 0.0     \\
game         & 100.0     & 97.9   & 98.9    & dislikeness      & 15.8      & 60.0   & 25.0    \\
hue\_lightchange & 85.7  & 93.3   & 89.4    & volume\_other    & 0.0       & 0.0    & 0.0    \\
querycontact & 89.8      & 100.0  & 94.6    & greet            & 9.5       & 80.0   & 17.0    \\
 \bottomrule
\end{tabular}}
\label{tab:results_english2}
\end{table}

\begin{figure}[!ht]
        \setlength{\tabcolsep}{0pt} % Default value: 6pt
    \renewcommand{\arraystretch}{0} % Default value: 1
    \centering
    \begin{tabular}{cc}
    \includegraphics[width=0.5\linewidth]{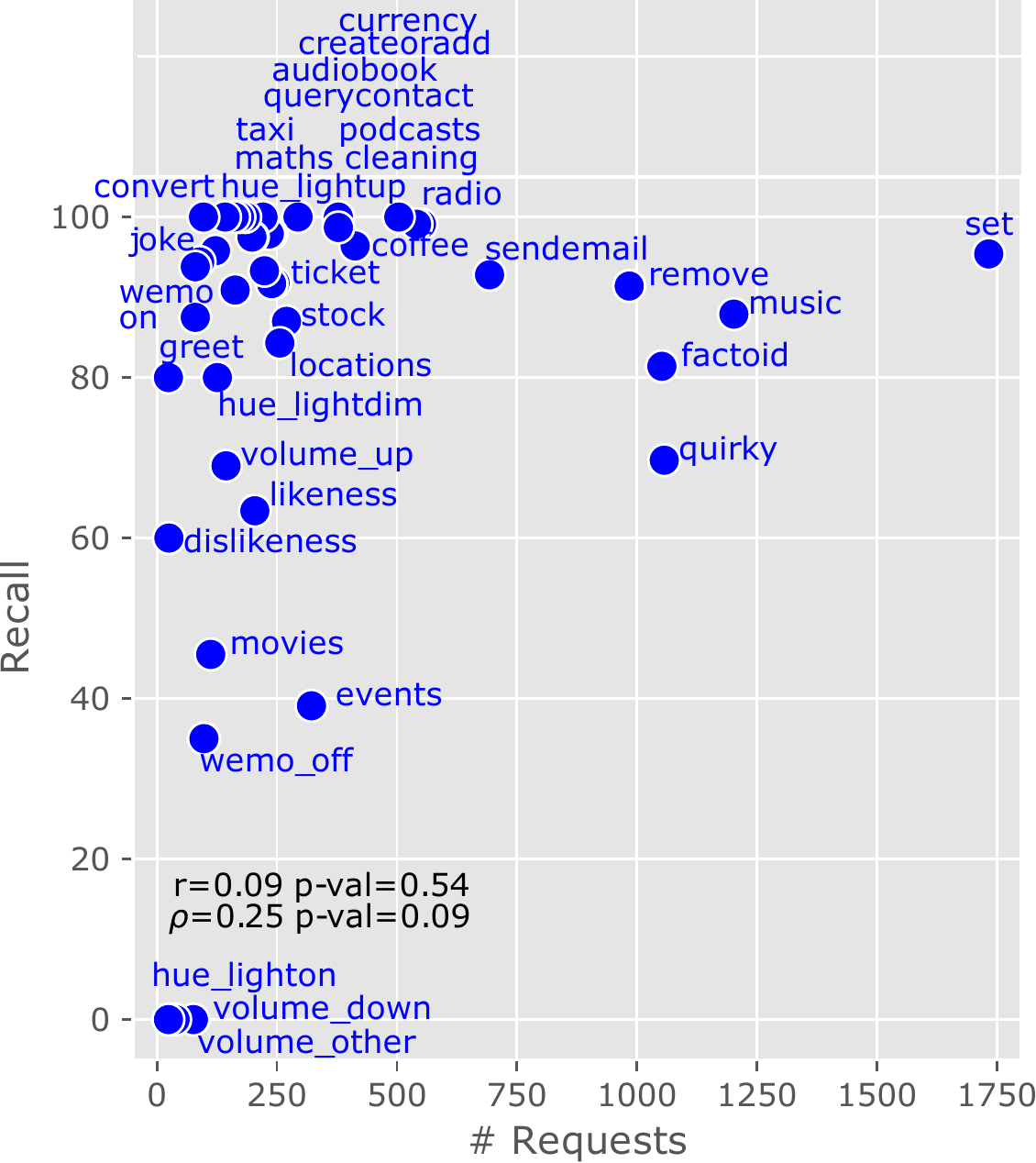}     &
    \includegraphics[width=0.5\linewidth]{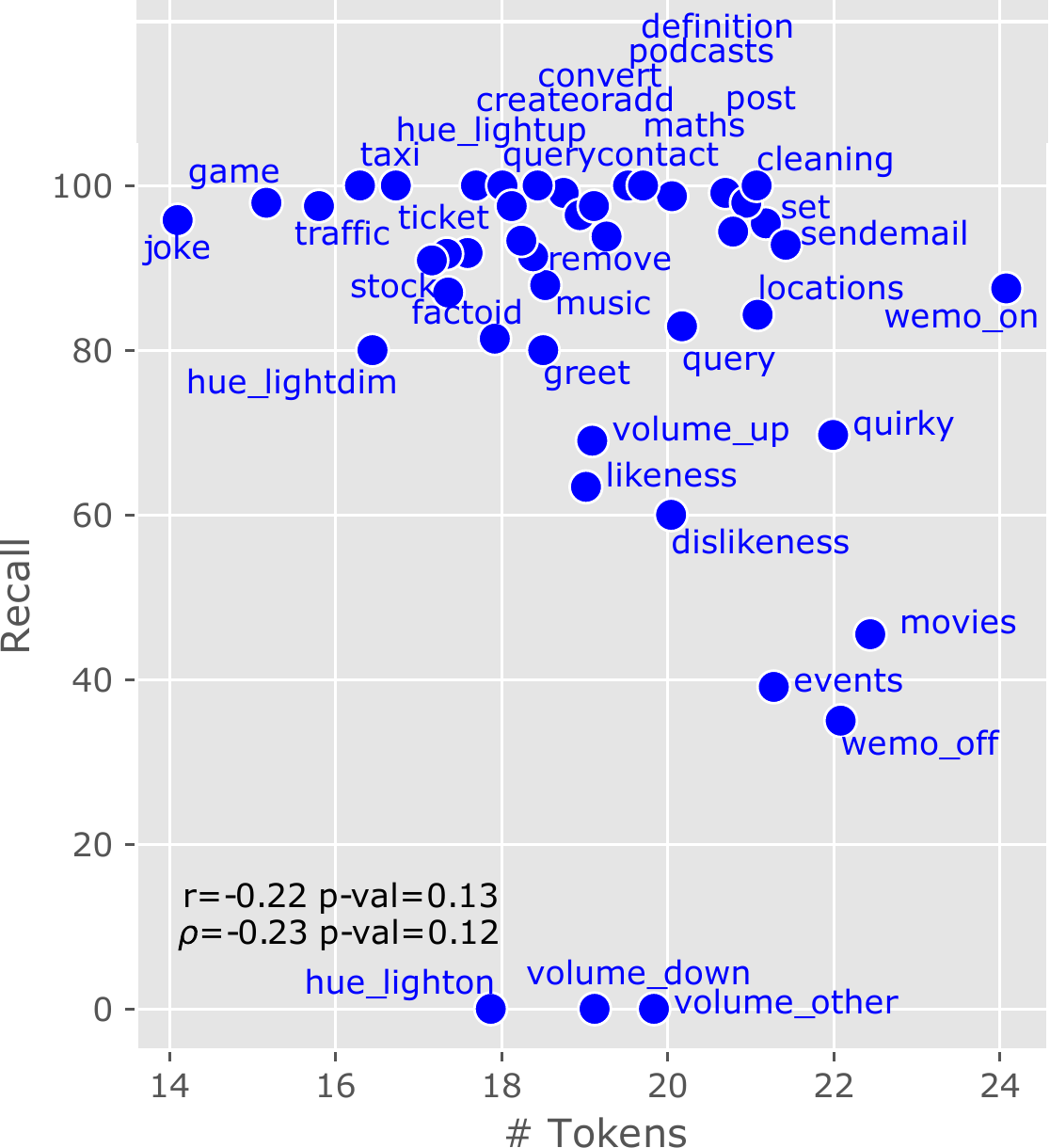}
    \end{tabular}
    \caption{Correlation between recall obtained for each intent in the in the English dataset and: \textbf{a)} number of requests for each intent. \textbf{b)} numbers of tokens for each intent.}
    \label{fig:analysis_english}
\end{figure}

The recall obtained for each intent is also compared against the number of request that are available for each intent (see Figure~\ref{fig:analysis_english}a)) and the average number of tokens to represent each intent (see Figure~\ref{fig:analysis_english}b)). The Pearson's ($r$) and Spearman's ($\rho$) correlation coefficients are also computed. For this case, there is no significant correlations between the number of requests and the recall obtained for each intent. However, this could be attributed to the number of request per intent for the English corpus is much lower than for the case of the SC$^2$ dataset. Higher correlation coefficients can be obtained if more samples are available for each intent. The results in Figure~\ref{fig:analysis_english}b) indicate that there is no correlation between the number of tokens and the recall observed for each intent, in the same way as the observed for the SC$^2$ dataset.

\section{Discussion}

We considered and compared the performance of different word embedding configurations based on word2vec and BERT in combination with several classification models based on convolutional and recurrent neural networks for intent recognition in multiple customer service chatbots. For the SC$^2$ dataset, which covers intents that are universal to multiple consumer service chatbots, we found that the most specific word embedding model (word2vec embeddings trained directly with the intent's data) achieved the highest accuracies. The most general word embedding model (BERT) shows to be the less accurate one. For the English corpus, which is based on intents related to human robot interaction, the most accurate models were also found using the word2vec embeddings, but using the pre-trained model from google news instead of the specific model trained directly with the chatbot's data. This is explained because the low number of words used to train the specific word2vec embedding with the chatbot data in English. While the w2v-sc2 embedding in Spanish is trained with 34,310 unique words, the vocabulary used to train the w2v-bot in English is formed only with 311 unique words. 

Regarding the comparison of different classification models, the most accurate model for the Spanish corpus corresponds to the CNN with parallel convolutions, while for the English data, the best result was obtained using the simple CNN model. In both cases, there is no much difference in the accuracy of the models, regarding the classification network.

We particularly observed a high reduction in the accuracy for those intents in the SC$^2$ dataset that are under-represented. Examples of under-sampled and inaccurate intents include asking for help and leaving negative comments. There is a significant Spearman's correlation ($\rho=0.77$) between the recall observed for each intent and the number of samples available in the corpus. A similar behavior is observed in the English corpus, however, in such a case the correlation is lower and non-significant. In general, in order to improve the accuracy of the considered models, the number of samples of the under-represented classes should be highly increased. The results obtained in both corpora also suggest that the accuracy to recognize each intent is independent on the number of tokens in which the intent is requested. The observed accuracy is equally distributed both for short and long intent's requests.

Finally, results from the confusion matrix in the SC$^2$ dataset suggested that most errors occur classifying intents with similar semantic fields in the dataset such as the misclassification between farewells and positive comments. In many cases, requests to the chatbot related to acknowledge for the provided service at the end of the conversation can be classified either as farewell or positive comments.  A similar behavior can occur when the user is not satisfied with the service and request to talk with a human agent, to get information about the attention line, or in general to make a negative comment to the chatbot.

\section{Conclusion}

We propose the creation of an intent recognition system able to detect several intents that are universal to several customer service chatbot systems. The set of universal intents include salutation, switch to a human agent, farewell, PCCs, among others. The core idea of the proposed system is to avoid repetitive training process when designing customer service chatbot systems, and at the same time to serve as a base model to design specific chatbot systems for different industries like finance, health-care, transport, insurance, among others. If these universal intents are accurately recognized, the training process of specific service chatbots becomes more straightforward because the person in charge of training the bot does not have to consider the training process of such intents.

The proposed system is based on the computation of word-embeddings combined with a deep learning classification model. We considered the computation of both context-independent (word2vec) and context-dependent (BERT) word-embedding models, which are classified using several neural network architectures based on convolutional and recurrent layers. 

The results indicated that it is better to consider context-independent models such as word2vec because the reduced and specific vocabulary, which is used for this type of applications. Conversely, models such as BERT, which are designed for wider and more natural language conversations produce lower accuracies for the considered problem.
The results also indicated that there is no high differences in the accuracy of the considered classification models. However, simpler models such a traditional CNN, or a parallel CNN produces slightly higher accuracies.

For future studies, the proposed intent recognition system can be used as well as a base model to train chatbot systems with dedicated intents, which are specific for certain industries such as finance, health-care, transport, retail,  among others. The proposed systems can be fine-tuned using transfer learning approaches to design better and more robust specific chatbot systems. 

\section*{Acknowledgment}
This work has been  funded by Pratech Group.

\bibliography{mybibfile}

\end{document}